\documentclass[showpacs,showkeys,nofootinbib,amsmath,amssymb]{revtex4-1}

\usepackage{hyperref}

\newcommand{\GG}{{\cal{G}}}
\newcommand{\VV}{{\cal{V}}}

\newcommand{\ud}{\mathrm{d}}

\newcommand{\bea}{\begin{eqnarray}}
\newcommand{\beal}[1]{\begin{eqnarray}\label{#1}}
\newcommand{\eea}{\end{eqnarray}} 
\newcommand{\be}{\begin{equation}} 
\newcommand{\bel}[1]{\begin{equation}\label{#1}}
\newcommand{\ee}{\end{equation}} 
\newcommand{\rf}[1]{(\ref{#1})}
\newcommand{\nn}{\nonumber}
\newcommand{\bit}{\begin{itemize}}
\newcommand{\eit}{\end{itemize}}
\newcommand{\ben}{\begin{enumerate}}
\newcommand{\een}{\end{enumerate}}

\newcommand{\brk}[1]{\left [ #1 \right ]}

\newcommand{\tq}{\tilde{q}}

\newcommand{\Lie}{\mathcal{L}}
\newcommand{\DD}{\mathcal{D}}

\newcommand{\tl}{\theta_{(\ell)}}
\newcommand{\tn}{\theta_{(n)}}
\newcommand{\tv}{\theta_{(v)}}

\def\half{\frac{1}{2}}

\def\tn{\theta_{(n)}}
\def\tl{\theta_{(\ell)}}

\def\Bn{{\cal B}}

\def\alp{\leavevmode\ifmmode {\alpha^\prime} \else ${\alpha^\prime}$ \fi}
\def\GN{G_{N}}

\begin{document}

\title{On the apparent horizon in fluid-gravity duality}

\author{Ivan Booth}
\email[]{ibooth@mun.ca}
\affiliation{Department of Mathematics and Statistics\\
Memorial University of Newfoundland\\
St. John's, Newfoundland and Labrador, A1C 5S7, Canada}

\author{Michal P.~Heller}
\email[]{m.p.heller@uva.nl}
\altaffiliation[On leave from: ]{\it So{\l}tan Institute for Nuclear Studies,
  Ho{\.z}a 69, 00-681 Warsaw, Poland} 
\affiliation{
\it Instituut voor Theoretische Fysica, Universiteit van Amsterdam\\ 
\it Science Park 904, 1090 GL Amsterdam, The Netherlands}

\author{Grzegorz Plewa}
\email[]{g.plewa@ipj.gov.pl}
\affiliation{So\l tan Institute for Nuclear Studies, Ho{\.z}a 69, 
00-681 Warsaw, Poland}

\author{Micha\l\ Spali\'nski}
\email[]{michal.spalinski@fuw.edu.pl}
\affiliation{
\it So{\l}tan Institute for Nuclear Studies, Ho{\.z}a 69, 00-681 Warsaw,
Poland \\ 
\it and Physics Department, University of Bialystok, 15-424 Bialystok,
Poland.}
 
\begin{abstract}

This article develops a computational framework for determining the location
of boundary-covariant apparent
horizons in the geometry of conformal fluid-gravity duality in arbitrary
dimensions. In particular, it is shown up to second order and conjectured to
hold to all orders in the gradient expansion that there is a unique apparent
horizon which is covariantly expressible in terms of fluid velocity,
temperature and boundary metric. This leads to the first explicit example of
an entropy current defined by an apparent horizon and opens the possibility
that in the near-equilibrium regime there is preferred foliation of apparent
horizons for black holes in asymptotically-AdS spacetimes.

\end{abstract}

\pacs{11.25.Tq,
04.50.Gh,
04.20.Gz.
}

\keywords{Gauge/gravity duality, Black Holes, Quasilocal horizons.}

\maketitle

\section{Introduction}

The AdS/CFT correspondence
\cite{Maldacena:1997re,Witten:1998qj,Gubser:1998bc}, or more generally
gauge-gravity duality, demonstrates a deep and fascinating link between black
hole physics and the plasma phase dynamics of certain (holographic) strongly
coupled gauge theories.  Over the last 10 years the gravitational side of the
correspondence has also started to be used as a tractable theoretical model of
strongly interacting non-Abelian media with properties similar to quark-gluon
plasma studied first at RHIC and now also at the LHC (see
\cite{CasalderreySolana:2011us} for the most recent review of these
developments). Early efforts in the applications of gauge-gravity duality
methods to hot QCD matter were motivated by hydrodynamic simulations of the
expanding fireball created in heavy ion collisions and focused on obtaining
transport properties of holographic plasmas by analyzing low-lying quasinormal
modes and linear response theory. These results provided concrete numerical
predictions for the simplest transport coefficients of strongly coupled
non-Abelian media with $\mathcal{N} = 4$ super Yang-Mills plasma as the
primary example\footnote{For an excellent review of these early developments
  see \cite{Son:2007vk}.} and have eventually led to the formulation of
fluid-gravity duality \cite{Bhattacharyya:2008jc}. Fluid-gravity duality is a correspondence
which maps solutions of relativistic Navier-Stokes equations describing
holographic liquids to long-wavelength distortions of black branes in higher
dimensional geometry. The direct connection between the dynamics of black
objects in higher dimensional spacetimes and solutions of nonlinear
hydrodynamics provided an opportunity to understand black brane geometries
and their features in terms of dual fluids, as well as to gain insights about
hydrodynamics from the properties of Einstein's equations. 
These perspectives,
as well as the possibility of applications, have generated significant interest
in the nonlinear dynamics of black brane spacetimes.

Dynamical black holes and their characterization has also been an important
research theme in mathematical relativity for the last couple of decades
(see \cite{Ashtekar:2004cn, Booth:2005qc, Gourgoulhon:2005ng} for useful reviews of this subject). The 
exact characterization of a dynamical black hole has proven to be a
surprisingly thorny theoretical problem for general relativity.
 The standard textbook definition associates black hole
interiors with regions of spacetime from which no signal can ever escape
\cite{Hawking:1973uf}. Thus, finding the exact extent of such a region is
necessarily a teleological procedure: properly defining ``ever'' and
``escape'' means that one must examine the ultimate fate of all signals from a
point before ruling whether or not that point is part of the black
hole. Identifying the event horizon boundary of a causal black hole is
similarly teleological. Thus, even though an event horizon is a congruence of
null geodesics obeying the same rules as any other congruence, its evolution
can appear to be acausal. For example the area increase of an event horizon is
not directly driven by infalling matter or energy; instead the actual effect
of an influx through the event horizon is a decrease in its rate of
expansion. 

These observations are not just mathematical curiosities. The non-local nature
of the event horizon is acceptable as long as one treats it as a causal
boundary removing the region containing a curvature singularity from the
dynamics of the rest of spacetime\footnote{Hence guaranteeing consistency of
  low energy description of black holes in terms of classical gravity.} and
does not associate any physical characteristics with it. However, this is not
the only role of the event horizon -- for the last four decades, one of the
most celebrated results of black hole physics has been the link between the
area of the event horizon and entropy. Already in the 1960s it was established
that event horizons necessarily increase in area \cite{Hawking:1973uf} and
this has usually been interpreted as being equivalent to the second law of
thermodynamics. Thus, the apparently acausal expansion of event horizons would
seem to imply a similarly acausal evolution of entropy. This leads to problems
since the origin of black hole entropy needs to be sought within microscopic
theories underlying gravitational interactions in the sense of the holographic
principle \cite{'tHooft:1993gx,Susskind:1994vu}. For asymptotically flat or
asymptotically de Sitter spacetimes such theories are not known (in principle
one could imagine any of these being non-local \cite{Li:2010dr}), but in the
anti-de Sitter context these are local quantum field theories in a suitably
understood large $N_c$ limit. At the superficial level it is hard to judge
whether acausality of the event horizon is a real problem in the first two
cases, 
but in the AdS/CFT context it definitely is.

The teleological nature of the event horizon is one of the main motivations
for the ongoing research program to characterize black holes (quasi)locally,
identifying their interiors from  the presence of strong gravitational fields
rather than on the basis of the causal structure of the entire spacetime.
Quasilocal horizons go by such names as trapping \cite{Hayward:1993wb}, isolated or 
dynamical horizons (the last two are both reviewed in \cite{Ashtekar:2004cn}). In 
all cases though, these horizons can be thought of as generalizations of the classical 
apparent horizons \cite{Hawking:1973uf}. Recall that apparent horizons are
associated with foliations of spacetimes. Areas of strong gravitational field 
are identified with the region on each surface that is covered by trapped surfaces. 
The boundary of that region is an apparent horizon and it is the outermost surface
for which the outgoing light front does not expand in
area\footnote{It is also required that ingoing light front shrinks in area and
  that inside this surface there are other surfaces, such that both ingoing
  and outgoing light fronts emitted from them shrink in area. For a more
  precise definition see Section \ref{quasilocalhorizons} or the review articles
  \cite{Booth:2005qc, Ashtekar:2004cn}.}. With a slight abuse of terminology
the union of such surfaces over all time slices is often also referred to as an
apparent horizon and it is in this sense that it will be used it here.

In the context of gauge-gravity duality characteristics of black holes
with planar horizons in higher dimensional spacetimes are at the same time the
quantities describing dual finite energy density or finite charge density
states of local quantum field theories. In static situations, where the
acausal nature of the event horizon plays no role, the entropy defined by the
event horizon was identified with the thermodynamic entropy of a dual
holographic field theory. Such entropy satisfies a very strong constraint: the
first law of thermodynamics linking IR quantities (i.e. temperature and
entropy density) with UV quantities (the energy density). Since the latter are
well defined in the dual quantum field theories (energy density is the
expectation value of one of the components of the energy-momentum tensor in
thermal equilibrium), there are no controversies with associating
thermodynamic entropy with the event horizon in time-independent
situations. However in static situations (at least in the context of
Kerr-Newman black holes/branes) the event horizon and one of apparent horizons
coincide, so that by associating the entropy with the event horizon one
actually associates it at the same time with an apparent horizon. The latter
identification actually turns out to be more robust, as the example of
conformal soliton flow \cite{Friess:2006kw} suggests \cite{Figueras:2009iu}.

Beyond equilibrium three problems arise. The first follows from the
aforementioned nonlocality of the event horizon, the second comes from
foliation dependence of apparent horizons, whereas the third one is related to
the various ways in which one can associate points on any of horizons with
points on the boundary (this is referred to as the bulk-boundary map
\cite{Bhattacharyya:2008xc}). The last of these is necessary to localize the
entropy production in the dual field theory. To illustrate that the first
issue is a serious problem quite disconnected from any ambiguities of the
bulk-boundary map, one can consider the example of gravitational dynamics with
a sharp distinction between a dual equilibrium regime without entropy
production and a dynamical transition with dissipation. The relevant
backgrounds, much in the spirit of the Vaidya solution, appeared in the
context of the thermalization problem of strongly coupled non-Abelian media
and describe gravitational processes in which the dual quantum field theory
undergoes a transition between two equilibrium states in a finite time
interval \cite{Chesler:2008hg,Chesler:2009cy}. In such a situation the event
horizon evolves past the bulk lightcones spanned by the transition region on
the boundary. The latter patch of bulk spacetime is dual to the boundary
region where the holographic field theory is in equilibrium, so that its
thermodynamic entropy stays constant. This result strongly suggests that the
causal boundary of a black hole is not the relevant entropy carrier regardless
of any ambiguities of the bulk-boundary map \cite{Figueras:2009iu}. In
contrast with the area of the event horizon, the entropy defined by the unique
apparent horizon respecting symmetries of 1-dimensional boundary dynamics
considered in \cite{Chesler:2008hg,Chesler:2009cy} was constant before the
transition process and eventually in the far future agreed with the one given
by the event horizon.

In less symmetric situations, apart from the choice of bulk-boundary map, the
foliation dependence of apparent horizons becomes a significant issue --
different foliations of spacetime lead to different apparent horizons. The
most trivial and at the same time pessimistic possibility is that the notion
of local entropy does not extend beyond equilibrium situations and foliation
dependence, as well as the freedom of bulk-boundary mapping, signal exactly
this. It might also be that on the dual field theory side there are many
relevant local notions of entropy and different apparent horizons correspond
to such different notions. Yet another possibility is that the field theory
notion of entropy does not suffer from ambiguities of kinds introduced by
foliation dependence of apparent horizons, which might be used as a guiding
principle for finding preferred apparent horizon.

Resolving these issues in the general case is very difficult if not
impossible, so the only hope is to proceed example by example. In global
equilibrium the foliation dependence essentially vanishes and the event and
apparent horizons coincide. In the near-equilibrium regime one expects the
horizons to be ``close'' (in a sense of \cite{Nielsen:2010h} or
\cite{Booth:2010eu}) so that a dynamical apparent horizon behaves almost like
an isolated horizon. If one takes the near-equilibrium regime as point of
departure for further studies, one is immediately led to considering apparent
horizons in the geometry of fluid-gravity duality. This background captures
the hydrodynamic regime on the field theory side starting from a locally
boosted and dilated black brane supplemented with gradient corrections
\cite{Bhattacharyya:2008jc} and from this perspective hydrodynamics can be
regarded as the simplest (because of its universality) type of collective dynamics that
quantum field theory can undergo.

The generalization of entropy to hydrodynamics is provided by the notion of an
entropy current. Such a current is constructed phenomenologically in the
gradient expansion by requiring that in equilibrium it reproduces
thermodynamic entropy and that its divergence evaluated on solutions of the
equations of hydrodynamics is non-negative. A detailed analysis of the
consequences of this generalized second law of thermodynamics on the form of
the entropy current in \cite{Romatschke:2009im} showed that even up to second
order in gradients there is an ambiguity inherent in such a
definition\footnote{Part of the ambiguity is trivial and comes from a term
  whose divergence vanishes.}. From the point of view of fluid-gravity duality
it was very natural to ask what is the gravity interpretation of the
coefficients appearing in the boundary hydrodynamic entropy current and what
might be the bulk counterpart of the ambiguity in its definition. In the
pioneering work \cite{Bhattacharyya:2008xc} a candidate entropy current was
obtained by mapping the area theorem on the event horizon onto the boundary
along ingoing null geodesics. In general, there are infinitely many directions
in which such geodesics can propagate from the boundary, but hydrodynamic
covariance requires that such geodesics close to the boundary move in the
direction specified (at leading order of the gradient expansion) by the local
fluid velocity \cite{Bhattacharyya:2008xc}. Ambiguities appearing in such
procedure appear at third and higher orders in the gradient expansion and were
irrelevant in the second order construction of
\cite{Bhattacharyya:2008xc,Romatschke:2009im}. This causal bulk-boundary map
can be supplemented with suitably understood boundary diffeomorphisms and the
latter turn out to capture precisely the ambiguity discussed in
\cite{Romatschke:2009im}.

Furthermore, in a recent paper \cite{Booth:2009ct} it was shown that for a
fixed bulk-boundary map the same freedom in entropy current can be understood
as coming from different bulk hypersurfaces with a fixed foliation satisfying
a generalized area theorem and asymptoting to the event horizon. Such surfaces
were dubbed ``generalized horizons'' with the event horizon and the (asymptoting
to it) apparent horizon being just two particular instances of the more general
notion. From the perspective of the phenomenological definition of 
the hydrodynamic entropy current none of these hypersurfaces and none of the
available bulk-boundary maps is favored over any other. However, causality of
the boundary field theory seems to favor the entropy current dual to the
apparent horizon -- providing that it is free of the ambiguities related to
foliation dependence and that the bulk-boundary map in use is causal. The task of
this paper is to elaborate on the proposal \cite{Booth:2010kr} by presenting a
derivation of the apparent horizon in the geometry of fluid-gravity duality, 
its features, as well as discussing the properties of the dual entropy current.

The organization of the paper is the following. Section \ref{introhydro}
presents in a self-contained fashion the geometry dual to conformal fluid
dynamics in arbitrary dimensions obtained in
\cite{Bhattacharyya:2008mz}. Section \ref{quasilocalhorizons} is the main part
of the paper and provides the detailed calculation of the relevant apparent
horizon in the case of conformal fluid-gravity duality up to second order in
gradients. Section \ref{SECTIONhydrocurrent} focuses on the hydrodynamic side of
fluid-gravity duality and analyzes the dual entropy current using the 
technology introduced in \cite{Booth:2010kr}. The general discussions of the
results and possible future directions of research are provided in Section
\ref{conclusions}. Appendix \ref{appendixA} provides some details on the 
Weyl-covariant derivative and Weyl-covariant hydrodynamic tensors, whereas
Appendix \ref{appendixB} illustrates the methods developed in Section
\ref{quasilocalhorizons} by describing the construction of an apparent horizon
in the Vaidya spacetime. Readers interested mostly in the 
general-relativistic aspects of these considerations can skip Section
\ref{SECTIONhydrocurrent} and regard the paper as an example of a perturbative
calculation of an apparent horizon in a geometry governed by 
Einstein's equations with negative cosmological constant.

\section{The geometry of fluid-gravity duality\label{introhydro}}

The geometry of fluid-gravity duality in arbitrary dimensions
\cite{Bhattacharyya:2008mz} is a solution to Einstein gravity with negative
cosmological constant
\bel{universalgravityaction}
I_{d+1} = \frac{1}{16 \pi \GN} \int \ud^{d+1} x \, \sqrt{-G} \left\{ R +
d\left(d-1\right) \right\},
\ee
where $\GN$ is the $(d+1)$-dimensional Newton's constant and the AdS radius is set to 1. 
The action \eqref{universalgravityaction} arises in the context of string
theory (for $d = 2,3,4$ and $6$, see \cite{Haack:2008cp} for details) and
describes a sector of decoupled dynamics of the one-point function of the
energy-momentum tensor operator in planar strongly coupled holographic
conformal field theories \cite{Bhattacharyya:2008mz}. The equations of motion
derived from \eqref{universalgravityaction} support a 
$d$-parameter family of exact, static black hole solutions with planar horizons
obtained by boosting and dilating the AdS-Schwarzschild black brane solution 
\bel{boostedblackbrane}
\ud s^{2} = 2 u_{\mu} \ud x^{\mu} \ud r - r^{2} \left( 1 - \frac{1}{\left( r b
  \right)^d} \right) u_{\mu} u_{\nu} \ud x^{\mu} \ud x^{\nu} + r^{2} \left(
g_{\mu \nu} + u_{\mu} u_{\nu} \right) \ud x^{\mu} \ud x^{\nu} \, .
\ee
Here $g_{\mu \nu}$ denotes components of the flat Minkowski metric on the conformal
boundary of the asymptotically AdS spacetime \rf{boostedblackbrane}. 
The boost parameter $u^{\mu}$ is a $d$-component velocity in the $x^{\mu}$
directions, normalized so that 
$u_{\mu} u^{\mu} = -1$ in the sense of the boundary metric
$g_{\mu \nu}$. The lines of constant $x^{\mu}$ in \eqref{boostedblackbrane}
are ingoing null geodesic, for large $r$ propagating in the direction set by
$u^{\mu}$, and the radial coordinate $r$ parametrizes them in an affine way
\cite{Bhattacharyya:2008xc}. The geometry \eqref{boostedblackbrane} may be
regarded as a stack of constant-$r$ $d$-dimensional planes, starting from the
boundary at $r = \infty$ (which is $d$-dimensional Minkowski spacetime) right
down to the curvature singularity at $r = 0$. The latter is shielded by the
event horizon at $r = 1/b$, which is at the same time an (isolated) apparent
horizon.   
The dilation parameter $b$ appearing in \eqref{boostedblackbrane} is related
to the Hawking temperature $T$ of the event horizon by  
\be
b = \frac{d}{4 \pi T} \, .
\ee
Unlike black holes in asymptotically flat spacetime, the metric
\eqref{boostedblackbrane} supports perturbations varying much slower within
the transverse planes than within the radial direction. The parameter 
controlling the scale of variations in the radial
direction is $b$. 

If $b$, $u^{\mu}$ and $g_{\mu \nu}$ are allowed to vary
slowly compared to 
the scale set by $b$, the metric \eqref{boostedblackbrane} should be an
approximate solution of nonlinear Einstein's equations with corrections
organized in an 
expansion in the number of gradients in the $x^{\mu}$ directions. Direct
calculations  
\cite{Bhattacharyya:2008jc} have shown that proceeding in this way is a
systematic way of solving Einstein's equations, provided that the dual
energy-momentum tensor \cite{Bianchi:2001kw,deHaro:2000xn} depending on $b$
and $u^{\mu}$ is conserved. For the metric \eqref{boostedblackbrane} the dual 
energy-momentum tensor is that of a relativistic perfect fluid with a conformal
equation of state 
\bel{pf}
T^{\mu \nu} = \epsilon u^{\mu} u^{\nu}  + p P^{\mu \nu}  \ ,
\ee
where
\be
P_{\mu \nu} = g_{\mu \nu} + u_{\mu} u_{\nu}
\ee 
is the projector operator onto the space transverse to $u^{\mu}$, 
\be
\epsilon = (d - 1) \, p = (d - 1) \frac{1}{16 \pi \GN} b^{-d}
\ee
and $g_{\mu \nu}$ is some weakly curved metric in which the fluid lives. 

Both
the metric \eqref{boostedblackbrane} and the energy-momentum tensor \eqref{pf}
receive gradient corrections. The separation of scales mentioned earlier
implies that corrections to the 
metric \eqref{boostedblackbrane} will be tensorial quantities made of
$x^\mu$-derivatives of $b$, $u^{\mu}$ and $g_{\mu \nu}$ with scalar
functions of $r$ and $b$ as coefficients. The relevance of these terms is suppressed
by the 
number of gradients and for practical reasons the expansion is terminated at
the 2-derivative level. The tensorial quantities in question are scalars $S$,
transverse ($u_{\mu} V^{\mu} = 0$) vectors $V^{\mu}$ and transverse ($u^{\mu}
T_{\mu \nu} = 0$) traceless symmetric rank 2 tensors $T_{\mu \nu}$\footnote{It
  needs to be borne in mind, that all these quantities are also required to be
  independent when evaluated on lower order solutions of hydrodynamics.}. A
priori 
one should consider all possible terms, as was done originally in
\cite{Bhattacharyya:2008jc}. This task can however be greatly simplified by
utilizing the underlying conformal symmetry and seeking Weyl-invariant
solutions of Einstein's equations, i.e. solutions invariant under simultaneous
rescalings of 
\bel{Weylrescalings}
g_{\mu \nu} \rightarrow e^{- 2\phi} g_{\mu \nu}, \quad u^{\mu} \rightarrow
e^{\phi} u^{\mu}, \ \quad b \rightarrow e^{-\phi} b \quad \mathrm{and} \quad r
\rightarrow e^{\phi} r 
\ee
where $\phi$ depends on the coordinates $x^{\mu}$
\cite{Bhattacharyya:2008mz}. The leading 
order metric \rf{boostedblackbrane} is Weyl-invariant, but due to the presence
of $\ud r$ it  
does not retain its form at higher orders. It can however be written in a
manifestly Weyl-invariant form upon introducing a vector field
$\mathcal{A}_{\nu}$ defined by \cite{Loganayagam:2008is}
\bel{connectionFIELD}
\mathcal{A}_{\nu} \equiv u^\lambda\nabla_\lambda u_{\nu}-
\frac{\nabla_\lambda  u^\lambda}{d-1} u_{\nu} \, .
\ee
This quantity is of order one in the gradient expansion and transforms as a
connection under Weyl-transformations 
\bel{coonectionFIELDtrafo}
\mathcal{A}_{\nu} \rightarrow \mathcal{A}_{\nu} +\partial_{\nu}\phi \, .
\ee
The Weyl-invariant form of the metric \eqref{boostedblackbrane} reads
\bel{boostedblackbraneWI}
\ud s^{2} = 2 u_{\mu} \ud x^{\mu} \left(\ud r - \mathcal{A}_{\nu} \ud
x^{\nu}\right) - r^{2} \left( 1 - \frac{1}{\left( r b \right)^d} \right)
u_{\mu} u_{\nu} \ud x^{\mu} \ud x^{\nu} + r^{2} \left( g_{\mu \nu} + u_{\mu}
u_{\nu} \right) \ud x^{\mu} \ud x^{\nu}. 
\ee
This metric is a leading order approximation to a spacetime whose metric is of
the form 
\bel{wimetric}
\ud s^2 = G_{ab} \ud x^a \ud x^b = -2 u_\mu \ud x^\mu (\ud r + \VV_\alpha \ud
x^\alpha) + \GG_{\mu\nu}  \ud x^\mu \ud x^\nu 
\ee
with the condition $u^{\mu} \GG_{\mu \nu} = 0$  completely fixing the gauge
freedom. The subleading corrections to \eqref{boostedblackbraneWI} need to be
Weyl-invariant and the simplest way to construct them is by summing individual
Weyl-invariant contributions order by order in the gradient expansion. A
single Weyl-invariant contribution to \eqref{wimetric} can be represented as a
scalar function of the Weyl-invariant combination $r b$ multiplying a
Weyl-covariant (i.e. transforming homogeneously under Weyl transformations of
$b$, $u^{\mu}$ and $g_{\mu \nu}$, see Appendix \ref{appendixA}) tensor
of a given weight $w$ supplemented with a factor of $b^{w}$.

A powerful tool in generating Weyl-covariant gradient terms is the
Weyl-covariant derivative $\mathcal{D}_{\mu}$, which uses the connection
$\mathcal{A}_{\mu}$ \eqref{connectionFIELD} to compensate for derivatives of
the Weyl factor coming from derivatives of Weyl-covariant tensors. It has the
property that a Weyl-covariant derivative of a Weyl-covariant expression is
itself Weyl-covariant with the same weight (see Appendix \ref{appendixA} or
the original publications 
\cite{Loganayagam:2008is,Bhattacharyya:2008xc,Bhattacharyya:2008mz} for
details).  

At first order in gradients there is only a single Weyl-covariant term
available, which is the shear tensor of the fluid $\sigma_{\mu \nu}$. It reads 
\bel{shear}
\sigma_{\mu \nu} = \frac{1}{2}\left(\DD_{\mu} u_{\nu} + \DD_{\nu} u_{\mu}\right)
\ee
and transforms with Weyl-weight $3$. At second order in gradients, there
are in total 10 Weyl-covariant terms: 3 scalars, 2 transverse vectors and 5
transverse traceless symmetric rank 2 tensors. For convenience these objects
can be defined with appropriate powers of $b$ to render them Weyl-invariant. 
The scalar contributions read 
\bel{scalars}
S_1 = b^2 \sigma_{\mu \nu} \sigma^{\mu \nu}, \quad
S_2  = b^2 \omega_{\mu \nu} \omega^{\mu \nu} \quad \mathrm{and} \quad
S_3  =  b^2 {\cal R},
\ee
where $\omega$ is the vorticity of the flow and ${\cal R}$ is the Weyl-covariant
curvature tensor and curvature scalar (see Appendix \ref{appendixA} for
details). The Weyl-invariant transverse vectors are  
\bel{vectors}
V_{1\, \mu}  = b P_{\mu \nu} \DD_\rho \sigma^{\nu \rho} \quad \mathrm{and}
\quad V_{2\, \mu}  = b P_{\mu \nu} \DD_\rho \omega^{\nu \rho} 
\ee
Finally, the Weyl-invariant tensors read 
\beal{tensors}
T_{1\, \mu \nu} = u^\rho \DD_\rho \sigma_{\mu \nu}, \quad
T_{2\, \mu \nu} = C_{\mu  \alpha  \nu \beta} u^\alpha u^\beta, \quad
T_{3\, \mu \nu} = \omega_\mu^\rho \sigma_{\rho \nu} + \omega_\nu^\rho \sigma_{\rho \mu}, \nonumber\\
T_{4\, \mu \nu} = \sigma_\mu^\rho \sigma_{\rho \nu} - \frac{1}{d-1} P_{\mu \nu}
\sigma_{\alpha \beta} \sigma^{\alpha \beta} \quad \mathrm{and} \quad
T_{5\, \mu \nu} = \omega_\mu^\rho \omega_{\rho \nu} + \frac{1}{d-1} P_{\mu \nu}
\omega_{\alpha \beta} \omega^{\alpha \beta},
\eea
where $C_{\mu \alpha \nu \beta}$ is a Weyl-covariantized curvature tensor
(consult Appendix \ref{appendixA} for 
its detailed form).

The metric \eqref{wimetric} up to second order in gradients can be expressed
in terms of \rf{shear}, \eqref{scalars}, \eqref{vectors} and
\eqref{tensors} and takes the form 
\beal{mc}
{\cal V}_{\mu} &=& r^2 \Bn u_{\mu} + r A_{\mu} \nonumber \\
     &+&
       \frac{1}{d-2} (- b^{-1} (V_{2\mu} + V_{1\mu}) + b^{-2} u_\mu ( S_2 - S_1
       + \frac{1}{2(d-1)} S_3) ) -
\frac{2 r}{(b r)^{d-1}} L V_{1\mu} +  \nonumber \\
       &+& u_\mu (\frac{1}{4} b^{-d-2} r^{-d} S_2 + \frac{1}{2 (d-1)}
       \frac{r^2}{(b r)^d} K_2 S_1)  \\ 
{\cal G}_{\mu \nu} &=& r^2 P_{\mu\nu} + 2 b r^2 F \sigma_{\mu\nu} + \nonumber \\
     &-& (T_{5\mu \nu} - \frac{1}{d-1} P_{\mu \nu} b^{-2} S_2) + 
     		    2 b^2 r^2 F^2 (T_{4\mu \nu} + \frac{1}{d-1} P_{\mu \nu} b^{-2} S_1) 
                    - \frac{2 r^2}{d-1} K_1 S_1 P_{\mu \nu} + \nonumber \\
                    &-& 2 b^2 r^2 H_1 (T_{1\mu \nu} + T_{4\mu
                    \nu} + T_{2\mu \nu}) + 2 b^2 r^2 H_2 (T_{1\mu \nu} + T_{3\mu
                      \nu}) 
\eea
where
\be
\Bn = - \frac{1}{2 (br)^d} (1 - (b r)^d).
\ee
The quantities $F,H_1,H_2,K_1,K_2,L$ are functions of $br$ introduced in
\cite{Bhattacharyya:2008mz} and read
\begin{equation}\label{metricfns:eq}
F(br)\equiv \int_{br}^{\infty}\frac{y^{d-1}-1}{y(y^{d}-1)}dy,
\end{equation}
\begin{equation}
H_1(br)\equiv \int_{br}^{\infty}\frac{y^{d-2}-1}{y(y^{d}-1)}dy,
\end{equation}
\begin{equation}
\begin{split}
H_2(br)&\equiv \int_{br}^{\infty}\frac{d\xi}{\xi(\xi^d-1)}
\int_{1}^{\xi}y^{d-3}dy \left[1+(d-1)y F(y) +2 y^{2} F'(y) \right]\\
&=\frac{1}{2} F(br)^2-\int_{br}^{\infty}\frac{d\xi}{\xi(\xi^d-1)}
\int_{1}^{\xi}\frac{y^{d-2}-1}{y(y^{d}-1)}dy,
\end{split}
\end{equation}
\begin{equation}
\label{K1br}
K_1(br) \equiv \int_{br}^{\infty}\frac{d\xi}{\xi^2}\int_{\xi}^{\infty}dy\ y^2 F'(y)^2,
\end{equation}
\begin{equation}
\label{K2br}
\begin{split}
K_2(br) &\equiv \int_{br}^{\infty}\frac{d\xi}{\xi^2}\left[1-\xi(\xi-1)F'(\xi)
  -2(d-1)\xi^{d-1} \right.\\ 
&\left. \quad +\left(2(d-1)\xi^d-(d-2)\right)\int_{\xi}^{\infty}dy\ y^2 F'(y)^2 \right],\\
\end{split}
\end{equation}
\begin{equation}
L(br)\equiv \int_{br}^\infty\xi^{d-1}d\xi\int_{\xi}^\infty dy\ \frac{y-1}{y^3(y^d
-1)}.
\end{equation} 
The metric given above is a solution of Einstein equations with negative
cosmological constant up to second order in gradients, provided that $b$ and
$u^{\mu}$ satisfy the equations of dual hydrodynamics, i.e. the equations of covariant
conservation of the energy-momentum tensor obtained from \rf{mc} by holographic
renormalization
\bel{enmomtensorUPtosecondORDER}
 T_{\mu \nu} = p (g_{\mu \nu} + d u_{\mu} u_{\nu}) - 2 \eta \sigma_{\mu \nu}
 - 2 \eta \tau_{\omega} ( T_{1 \, \mu \nu} + T_{3 \, \mu \nu} )+
 2 \eta b (T_{1 \, \mu \nu}+T_{2 \, \mu \nu}+T_{4 \, \mu \nu}),
 \ee
 where
 \be
 \eta = \frac{s}{4 \pi} = \frac{1}{16 \pi \GN  b^{d-1}}, \quad
 \tau_{\omega} = b \int_{1}^{\infty} 
 \frac{y^{d-2}-1}{y(y^d-1)} \textmd{d}y, \quad p = \frac{1}{16 \pi\GN} b^{d}.
 \ee
The geometric picture emerging is that of spacetime locally approximated by
tubes of uniform black branes spanned along ingoing null geodesics given by
lines of constant $x^{\mu}$. The dilation and boost parameters $b$ and
$u^{\mu}$, as well as the boundary metric $g_{\mu \nu}$ vary from tube to tube,
but, as anticipated, the scales of these variations are small compared to
variations of the bulk metric along the radial null direction
\cite{Bhattacharyya:2008xc}. Due to this tubewise approximation, the leading
order geometry of fluid-gravity duality inherits the causal structure of
static black brane, i.e. the event horizon located at $r = 1/b$, now with $b$
depending on $x^{\mu}$ \cite{Bhattacharyya:2008xc}. It is to be expected, and
is confirmed by direct calculation further in the text, that the event horizon
of \eqref{boostedblackbrane} with slowly varying $b$, $u^{\mu}$ and $g_{\mu
  \nu}$ is at the same time an apparent horizon. Such an apparent horizon is
called isolated and does not lead to entropy production. The isolated apparent
horizon at $r = 
1/b$ is expected to become dynamical once corrections to \eqref{boostedblackbrane}
are included and its position will also be modified. The dynamics of this
almost isolated apparent horizon can be described in a gradient expansion 
much in the spirit of the framework of slowly evolving horizons
\cite{Booth:2003ji, Booth:2006bn,Kavanagh:2006qe}.

\section{Locating the apparent horizon in the geometry of fluid-gravity
  duality \label{quasilocalhorizons}}   

This section is devoted to identifying an apparent horizon for the spacetimes
defined by the metric (\ref{wimetric}). This search will be based on two
criteria: 1) from the isolated horizon contained in the unperturbed geometry 
(\ref{boostedblackbrane}) it is natural to expect the apparent horizon to be
a perturbation of the hypersurface $r=1/b(x)$ and 2) to ensure compatibility
with the dual conformal fluid solution those perturbations are required to be
manifestly Weyl-invariant.

\subsection{Preliminaries}

Apparent horizons are defined in terms of trapped and marginally trapped
surfaces.
In both cases the term ``surface''
means a codimension-two hypersurface $\Omega$ embedded in a larger
spacetime. The normal space to such a surface is spanned at any point by a
pair of null vectors $\ell$ and $n$. The following considerations apply to spacetimes 
where it makes sense to specify that both of these are future-oriented and 
respectively outwards and inwards pointing. It is convenient to 
cross-normalize them so that $\ell \cdot n = -1$ (this leaves a degree of
scaling freedom).

The induced metric on $\Omega$ can be written as 
\begin{equation}
\tilde{q}_{ab} = g_{ab} + \ell_a n_b + \ell_b n_a \, , \label{qdef}
\end{equation}
while the outward and inward null expansions of $\Omega$ are
\begin{equation}
\tl = \tilde{q}^{ab} \nabla_a \ell_b = \Lie_\ell \log \sqrt{\tq} \; \;
\mbox{and} \; \;  
\tn = \tilde{q}^{ab} \nabla_a n_b = \Lie_n \log \sqrt{\tq}
\label{tltn}
\end{equation}
or, more generally, for an arbitrary normal vector $X = A \ell + B n$
\begin{equation}
\theta_{(X)} = A \tl + B \tn \, . 
\end{equation}

Now $\Omega$ is said to be \emph{outer trapped} if $\tl < 0$, \emph{trapped}
if $\tl < 0$ and $\tn < 0$ and \emph{untrapped} if $\tl = 0$ and $\tn < 0$. It
is \emph{outer marginally trapped} if $\tl = 0$ and \emph{marginally trapped}
if $\tl = 0$ and $\tn < 0$. Trapped surfaces are indicative of black hole
regions, with well-known theorems linking them to both singularities and the
existence of event horizons \cite{Hawking:1973uf}. As recalled in the
introduction they are also used to define \emph{apparent
  horizons} \cite{Hawking:1973uf}. Given a foliation of spacetime into
spacelike hypersurfaces $\Sigma_t$ (``instants'' of time) one can define the
total trapped region on each $\Sigma_t$ as the union of all the outer trapped
surfaces.  Then (up to some technicalities which will be ignored here) the
boundary of each of those regions $\Omega_t$ is outer marginally trapped and
known as the apparent horizon. A common abuse of terminology (adopted in the
following) also uses the term apparent horizon to refer to the hypersurface
defined by the evolving $\Omega_t$ (that is the union of the $\Omega_t$).

In practical calculations, this definition of an apparent horizon is not very
usable and instead one just searches directly for hypersurfaces foliated by
outer marginally trapped surfaces. This is common practice in numerical
relativity (see, for example, \cite{Gourgoulhon:2005ng} and references
therein).  More generally, the teleological nature of classical black holes
and their event horizons has lead many to search for a (quasi)local and
properly causal replacement. Horizons foliated by (outer) marginally trapped
surfaces which (hopefully) bound regions of trapped surfaces are the most
obvious and mathematically tractable candidates.

For example, the boundaries of stationary black holes (or branes) are taken to
be \emph{weakly isolated horizons}: codimension-one hypersurfaces that are
foliated by outer marginally trapped surfaces or \emph{isolated horizons} if
their extrinsic geometry is also invariant (see for example the discussions in
\cite{Ashtekar:2000hw, Ashtekar:2001jb} or review articles such as
\cite{Ashtekar:2004cn, Booth:2005qc, Gourgoulhon:2005ng}).  These are closely 
related (though more general than) Killing horizons and under many
circumstances do a good job of characterizing a stationary black hole boundary
without reference to causal structure or infinities. This is particularly so
if one adds extra conditions to ensure that there are fully trapped surfaces
``just inside'' the horizon. For Hayward's \cite{Hayward:1993wb} future outer
trapping horizons (FOTHs)  one assumes that
\begin{equation}
\tn < 0 \; \; \mbox{and} \; \;  \Lie_n \tl  < 0 \, . 
\label{cond}
\end{equation}
That is, the inward expansion is negative and under a small inwards
deformation the outward expansion also becomes negative.  The black branes
considered in this paper are examples of FOTHs.  

For the classical definition, it is clear that time-evolved apparent
horizons are foliation dependent: different foliations will sample a different
set of trapped surfaces and so give rise to a different ``time-evolved''
horizon. Alternatively, focusing on the time-evolved horizon itself, it can
be shown (see, for example \cite{Ashtekar:2005ez, Booth:2006bn}) that a
hypersurface foliated by outer marginally trapped surfaces is not rigid and
may be deformed while maintaining its properties.  The non-uniqueness of
apparent horizons has been explicitly demonstrated in several papers
\cite{Schnetter:2006, Nielsen:2010}.

\subsection{Finding the horizon: strategy}

Problems with uniqueness are somewhat alleviated in the present calculation by
the xxrequirement that perturbations of the horizon be manifestly 
Weyl covariant. Then, the time-evolved apparent horizon $\Delta$ should be
specified as the level set of a scalar function 
\bel{formofs} 
S(r,x) = b(x) r - g(x) \, ,
\ee
where $g(x)$ is a Weyl-invariant scalar defined by
\be
g(x) = g_1(x) + g_2(x) \ + \dots 
\ee
where $g_k$ denotes a linear combination of all Weyl-invariant scalars at
order $k$ in the gradient expansion. There are no Weyl-invariant scalars at
order 1, and 3 at order 2, so one expects to find
\bea
g_1(x) &=& 0 \nn \\
g_2(x) &=& h_1 S_1(x) + h_2 S_2(x) + h_3 S_3(x) \, , 
\eea
where the $S_i$ are the 3 independent Weyl-invariant scalars \rf{scalars} and
the constants $h_i$ will be 
determined by solving $\tl=0$ and the conditions (\ref{cond}). Once this is
done, the expression for the 
position of the apparent horizon will take the form
\bel{horpos}
r = r_H(x) \equiv \frac{1}{b} \left(1 + h_1 S_1(x) + h_2 S_2(x) + h_3 S_3(x) 
\right) \, . 
\ee
This is a strong constraint, but a reasonable one to impose in a
perturbative regime where physical considerations suggest that the horizon
should 
be given by a Weyl covariant structure. Testing these surfaces as
potential horizons means that one must consider their possible  
foliations and find out whether any of them satisfy $\tl = 0$ and the
conditions \rf{cond}. Again however, one can lean on the Weyl covariance 
to simplify the calculation. Specifically, the outer marginally trapped
surfaces of $\Delta$ will have their own (in $\Delta$) normal $v$. This vector
is required to be expressible as a sum of Weyl invariant terms and further that
it be surface forming 
\bel{frob}
v \wedge dv = 0 \, . 
\ee
Though this only really needs to apply on the horizon itself, it turns out to
be computationally much easier to check this condition for  
$v$ specified not only on the putative horizon but also in some
neighbourhood. Thus, in practice one should look for Weyl-covariant one-form fields   
that are surface forming in some neighbourhood of $\Delta$. 

Thus the search domain will not be arbitrarily large, but rather be restricted
to potential horizons and foliations that are essentially Weyl-covariant
perturbations of the unperturbed boosted black brane solution
\rf{boostedblackbrane}.  Marginally outer trapped surfaces are to be sought
among intersections of these classes.  It will be shown in the following that
for the geometry of fluid-gravity duality, up to second order in the gradient
expansion the conditions $\tl = 0$ and \rf{frob} determine $v$ (as well as the
$h_i$ in \rf{horpos}) uniquely.

\subsection{Finding the horizon: hypersurfaces and intersections}

The program outlined above can be implemented as follows. The normal covector
to a surface of the form \rf{formofs} is 
\bel{ds}
m = dS \, , 
\ee
which up to second order in the gradient expansion is
\bel{masform}
m = r \partial_\mu b\ dx^\mu +  b\ dr \, .
\ee
The function $g$ does not contribute above, since the leading term involves 
$\partial_\mu g_2$, which is of third order in gradients. 
It is convenient to write the normal in terms of the Weyl-covariant 
derivative, which acting on $b$ (Weyl weight $-1$) is 
\be
\DD_\mu b = \partial_\mu b - \mathcal{A}_\mu b \, .
\ee
One then has\footnote{Recall
  now that the geometry found in \cite{Bhattacharyya:2008mz} 
  satisfies Einstein equations                                      
provided that the equations of hydrodynamics are satisfied by the      
quantities $b$, $u^\mu$ (in terms of which also $\mathcal{A}_\mu$ is expressed). These 
equations imply \cite{Bhattacharyya:2008mz} that $\DD b$ is of the second order
in gradients.}
\be
m = r \left(\DD_\mu b + b \mathcal{A}_\mu\right) dx^\mu +  b\ dr \, .
\ee
Raising the index using the metric \rf{wimetric} one gets (using the formula
for the inverse given in \cite{Bhattacharyya:2008mz}), up to second
order  
\bea
m^{\mu} &=& b {u}^{\mu}+ b^{2} \left(
{V_1}^{\mu} ( - \frac{2}{d}  \frac{1}{ b r} + \frac{1}{(d - 2)} \frac{1}{(b
  r)^2} + \frac{2}{(b r)^d} L) + {V_2}^{\mu} \frac{1}{(d - 2)} \frac{1}{(b
  r)^2} \right) 
 \\
m^r &=&  2\, {\cal B} r{}^{2} b - {A}_{\mu}  b r {u}^{\mu} + \nonumber \\
&+& \frac{1}{b} \left((\frac{2}{(d - 2)} + \frac{1}{2}
\frac{1}{(br)^d}) S_2 
+  ( - \frac{2}{d(d - 1)} br - \frac{2}{d-2}  + \frac{1}{(d - 1)} \frac{1}{(b
  r)^{d-2}} K_2) S_1 +   
\frac{1}{(d - 2)} \frac{1}{(d - 1)} S_3 \right).
\eea 

Next one must consider potential foliations of $\Delta$.
As noted earlier, the foliation of the apparent horizon can be specified by a
vector field $v$ which is tangent to $\Delta$ and but otherwise normal to
the leaves. Given a parametrization $y^\mu$ of the horizon so that 
$S(r(y), x(y)) \equiv const$, this means that
\bel{orth}
\frac{\partial S}{\partial r} \frac{\partial r}{\partial y^\alpha} +
\frac{\partial S}{\partial x^\beta} \frac{\partial x^\beta}{\partial y^\alpha}
= 0 \, . 
\ee
In terms of this coordinate system the tangent vectors are, of course,
$\partial/\partial y^\alpha$ which push-forward into the 
full space time as 
\be
 \vec{e}_\mu =  \frac{\partial x^a}{\partial y^\mu} \frac{\partial}{\partial
   x^a}  
\ee
and applying \rf{orth} one finds
\be
 \vec{e}_\mu =  \frac{\partial x^\beta}{\partial y^\mu} \left\{
\frac{\partial}{\partial x^\beta} - 
\left(
\frac{\partial S}{\partial r}
\right)^{-1} 
\left(
\frac{\partial S}{\partial x^\beta}
\right)
 \frac{\partial}{\partial r} 
\right\} \, . 
\ee
By construction these vectors all satisfy $v \cdot m = 0$ (as they should). 

It is very convenient to choose the coordinates on the horizon $y^\mu = x^\mu$.
This should be reasonable as long as the horizon does not ``fold over''; this  
should be the case in this perturbative, gradient expansion limit. This choice
also has the advantage  
of making the bulk-boundary map trivial (as discussed in Section
\ref{SECTIONhydrocurrent}). Then the tangent vectors to $\Delta$  
can be written as
\be
 \vec{e}_\beta =  
\frac{\partial}{\partial x^\beta} - 
\left(
\frac{\partial S}{\partial r}
\right)^{-1} 
\left(
\frac{\partial S}{\partial x^\beta}
\right)
 \frac{\partial}{\partial r} 
\ee
and a general vector field tangent to the horizon is given by 
\be
v = v^\mu \vec{e}_\mu \, .
\ee
In terms of the coordinate basis in 
the bulk one then has    
\bel{genv}
v =  v^\beta 
\left\{
\frac{\partial}{\partial x^\beta} - 
\left(
\frac{\partial S}{\partial r}
\right)^{-1} 
\left(
\frac{\partial S}{\partial x^\beta}
\right)
 \frac{\partial}{\partial r} 
\right\} \, .
\ee

Requiring that the vector $v$ be Weyl covariant fixes (up to second order) 
\beal{vsecord}
v^\mu &=& b {u}^{\mu} +  \{{V_1}^{\mu} b^{2} c_1 +
{V_2}^{\mu} b^{2} c_2 + 
{u}^{\mu} (S_1 b e_1  + S_2 b e_2  + S_3 b e_3) \}  \\
v^r &=&  - r {A}_{\mu} b {u}^{\mu} + S_1 r 
\frac{2}{d(d-1)} \,  ,
\eea
where $c_1$, $c_2$,  $e_1$, $e_2$ and $e_3$ are some constants. 
It is computationally convenient to 
normalize $v$ so that 
\bel{mvnormal} 
m^2 + v^2 = 0  \, ,
\ee 
in which case the coefficients of the longitudinal terms in \rf{vsecord}
vanish 
\be
e_1 = e_2 = e_3 = 0 \, . 
\ee
The remaining coefficients ($c_1, c_2$) appearing in $v$ are also not 
arbitrary. As discussed earlier, to ensure that the vector $v$ defines a
foliation one 
has to impose the Frobenius condition \rf{frob}. There are two types of terms,
which turn out to be given by 
\bea
v_{[\mu} \partial_\nu v_{\rho]} &=& 0 \nn \\
v_{[r} \partial_\nu v_{\rho]} &=& 
d \left(c_1 + \frac{1}{d} - \frac{1}{d-2}\right) V_{1[\nu} u_{\mu]} + 
d \left(c_2 - \frac{1}{d-2}\right) V_{2[\nu} u_{\mu]} 
\eea
up to terms of higher order in the gradient expansion. This
determines the coefficients $c_1$ and $c_2$
\bea
c_1 &=& \frac{2}{d(d-2)} \nn \\
c_2 &=& \frac{1}{d-2} \, .
\eea
This way one finds that the foliation vector $v$ is completely
determined once
$\Delta$ is fixed
\bel{vres}
v^{\mu} = b u^{\mu} + \frac{1}{d-2} b^2 \left( \frac{2}{d}\ V_1^\mu + V_2^\mu
\right) \, .  
\ee
Since the Frobenius condition was imposed for the full spacetime (rather than
just on the horizon), this vector $v$ actually gives rise to foliation of the
full 
spacetime, at least in a neighborhood of the horizon.

It is interesting that one gets a unique result. It seems plausible that this
will also be the case at higher orders in the gradient expansion. To see this, note that
at a given order $k$, $v$ is entirely specified in terms of its $v^{\mu}$
components, and its $v^{r}$ component does not depend on the $k$-th order
contribution to $v^{\mu}$. 
In complete analogy with the second order, $v^{\mu}$ at order $k$ will be a
linear combination of all available transverse and longitudinal vectors. The
hypersurface $\Delta$ is specified 
as the level set of a scalar function $S(r,x)$ (see \rf{formofs}), which at
order $k$ contains all the available hydrodynamic scalars of order $k$. The
vector 
$m$ normal to $\Delta$ is defined by $m=dS$, so the construction does not
introduce any further coefficients to be determined. Now consider the
normalization 
condition \eqref{mvnormal} and expand the $v^{2}$ contribution 
\bel{normalALLorders}
\GG_{\mu\nu} v^{\mu} v^{\nu} - 2 u_{\mu} \VV_{\nu} v^{\mu} v^{\nu} - 2 u_{\mu}
v^{\mu} v^{r} + m^{2} = 0. 
\ee
In order to evaluate the $k$-th order contribution to \eqref{normalALLorders} from
$v^{\mu}$ it is sufficient to take the zeroth order metric. Note however that
since at leading order $v^{\mu}$ is proportional to $u^{\mu}$ and
$\GG_{\mu\nu}$ is transverse, the first term on the left hand side of
\eqref{normalALLorders} vanishes for all $r$. Since $v^{r}$ does not receive
corrections from the $k$-th order $v^{\mu}$, the only term which depends on this is
actually $u_{\mu} \VV_{\nu} v^{\mu} v^{\nu}$. But since $\VV_{\mu}$ is also
proportional to $u_{\mu}$ at leading order, the whole left hand side of
\eqref{normalALLorders} at order $k$ depends only on the longitudinal
contributions to $v^{\mu}$ at this order. If so, formula
\eqref{normalALLorders} fixes them uniquely and does not constrain the transverse
contributions. What is left at order $k$ are scalar contributions to $S(r,x)$ and
transverse contributions to $v^{\mu}$. But at a given order, transverse
and longitudinal quantities are independent, so the scalar
condition $\theta_l$ at order $k$ fixes all the contributions to $S(r,x)$. 
The transverse components of $v^{\mu}$, relevant for the foliation of
$\Delta$, are
likely to be fixed by the Frobenius condition \eqref{frob} in analogy with
what happens at second order. It can be checked that the contributions in
question will appear in the Frobenius conditions, but it seems difficult to
show that by 
choosing the transverse parts appropriately one can satisfy Frobenius
conditions at any order. One argument that this is indeed the case is that
such a 
condition must be satisfied for $v^{\mu}$ on the event horizon at arbitrary order and
in this case $v^{\mu}$ is fixed and given by $m^{\mu}$. It would certainly be 
interesting to make these statements more precise.

\subsection{Horizons}

Dynamical quasilocal horizons are spacelike and so $m$ should be
timelike and $v$ spacelike. Without loss  
of generality one can assume that $m$ is future pointing and $v$ is outward pointing. 
Then the null normals to the surfaces of constant $S$ and $v$ are 
\beal{evos}
v &=& \ell - C n \nn \\
m &=& \ell + C n \, ,
\eea
where the scalar $C$ is called the evolution parameter
\cite{Booth:2003ji,Booth:2006bn}. In this case  
\bel{cform}
C = \half v^2 \, .
\ee
The sign of the evolution parameter indicates whether $\Delta$ is
spacelike or timelike (or null if $C=0$). The signs of the coefficients in
\rf{evos} have been chosen to 
ensure that both $\ell$ and $n$ are future-pointing, and $\ell$ is
outward-pointing while $n$ is inward-pointing.

The null normals are then
\bea 
\ell^\mu &=& b {u}^{\mu} + \nonumber \\
  &+& \half b^2  \left(
{V_2}^{\mu} ( \frac{1}{(d - 2)} \frac{1}{(br)^2} +  c_2) + 
{V_1}^{\mu} ( - 2\, \frac{1}{d} \frac{1}{rb} + \frac{1}{(d - 2)} \frac{1}{(rb)^2} + 2
L \frac{1}{(rb)^d}   +  c_1)
\right) \\ 
\ell^r &=& - {A}_{\mu}  b r {u}^{\mu} + {\cal B} r^{2} b + \nonumber \\
 &+& b^{-1}  \left(
S_2 (\frac{1}{d - 2} + \frac{1}{4} \frac{1}{(br)^d}) + 
S_1 ( - \frac{1}{d - 2}  + \frac{1}{2} \frac{1}{(d-1)} \frac{1}{(b r)^{d-2}} K_2) + 
\frac{1}{2(d - 2)} \frac{1}{(d - 1)} S_3\right) 
\eea
and 
\bea
n^\mu &=&  \frac{1}{ 2 {\cal B} r^2} \left(
V_1^{\mu} (\frac{2}{d} \frac{1}{rb}  - \frac{1}{(d -2)} \frac{1}{(rb)^2} -2
\frac{1}{(r b)^d} L  + c_1) +  
{V_2}^{\mu} (- \frac{1}{(d - 2)} \frac{1}{(rb)^2}  + c_2) 
 \right) \\
n^r &=&  - \frac{1}{b} \, .
\eea

As a check on the results obtained so far one can determine the location of
the 
event horizon and compare with \cite{Bhattacharyya:2008mz}.
Using the explicit form of the evolution parameter $C$ (obtained from \rf{cform}) 
\be
C =  - {\cal B} b^{2} r^{2} + 
(\frac{1}{(d - 2)} - \frac{1}{2} \frac{1}{(d-1)} \frac{1}{(b r)^{d-2}} K_2 
+ \frac{2}{d (d - 1)} b r ) S_1 - 
(\frac{1}{(d - 2)} + \frac{1}{4 (r b)^{d}}) S_2 -
\half \frac{1}{(d - 2)} \frac{1}{(d - 1)} S_3 
\ee
it is easy to calculate where the event horizon is located. On general grounds
one expects a result of the form \rf{horpos}.
Solving $C(r_{EH}) = 0$ one finds \rf{horpos} with 
\beal{EH}
h_1^{(EH)} &=& \frac{2 (d^2 + d -4)}{d^2 (d-1) (d-2)} - \frac{1}{d(d-1)} K_2(1)
\nonumber \\
h_2^{(EH)} &=& - \frac{d + 2}{2 d (d-2)} \nonumber \\
h_3^{(EH)} &=&  -\frac{1}{d (d-1) (d-2)} \, ,
\eea
which matches the results of \cite{Bhattacharyya:2008mz}.  

Note that because the event horizon is null it must be the case that $v$ is
proportional to $m$. Using \rf{EH} and the explicit form of $m$ and $v$, it
may be checked directly that this is indeed the case.

To determine the position of the apparent horizon one needs to calculate the
null expansions from the forms  \rf{tltn} 
\bea
\tl = \tilde{q}^{ab} \nabla_a \ell_b \; \; \mbox{and} \; \; 
\tn = \tilde{q}^{ab} \nabla_a n_b \, ,
\eea
where the metric induced on the foliation slices is calculated from
\rf{qdef}. Using the results of the previous section one finds (up to
second order)\footnote{This computation is fairly lengthy.} 
\beal{thetares}
\tl &=& (d-1) \left( {\cal B} br + \frac{1}{br} \left(
S_1 ( - \frac{1}{(d - 2)} + \frac{1}{2(d-1)} \frac{1}{(b r)^{d-2}} K_2
-\frac{1}{d-1} {\cal B} b^2 r^3 K_1'(1)) + \nonumber 
\right. \right. \\
&+& \left. \left. S_2 ( - \frac{1}{d-1} {\cal B} + \frac{1}{(d - 2)} + \frac{1}{4}
\frac{1}{(br)^d} ) + \half \frac{1}{(d-1) (d - 2)} S_3 
\right) \right) \\
\tn &=& - \frac{d-1}{br} + 
 \frac{1}{br}  \left(
 \frac{1}{(br)^2} S_2 + r K_1' S_1\right) \, .
\eea
Note that the results are manifestly Weyl-invariant. In particular, there is
no correction at first order (as required by Weyl invariance).   
With these results in hand, it is straightforward to determine the location of
the apparent horizon by solving $\theta_{(\ell)}(r_{AH}) = 0$. One again finds
\rf{horpos} with 
\beal{AH}
h_1^{(AH)} &=& \frac{2}{(d-2) d} - \frac{1}{d(d-1)}  K_2(1), \nonumber \\
h_2^{(AH)} &=& - \frac{d + 2}{2 d (d-2)}, \nonumber\\
h_3^{(AH)} &=&  -\frac{1}{d (d-1) (d-2)} \, .
\eea
Only $h_1$ differs from the result for the event horizon \cite{Bhattacharyya:2008mz}. 

The expression 
\be
r_{EH} - r_{AH} = \frac{4 S_1}{b d^2 (d-1)} \geq 0
\ee
explicitly shows that the apparent horizon lies within (or coincides with) the
event horizon in the sense that an ingoing radial null geodesic will cross first the
event horizon and only then the apparent horizon, since $r$ is an affine
parameter on such geodesics. It is also easy to check that the apparent
horizon is spacelike or null 
\be
C(r_{AH}) = \frac{2 S_1}{d (d-1)} \geq 0 \, ,
\ee
as required.

\section{The hydrodynamic entropy current defined by the apparent
  horizon\label{SECTIONhydrocurrent}}  

In hydrodynamics the entropy current is a phenomenological notion constructed
order-by-order in the gradient expansion starting from the term describing the
flow of thermodynamic entropy. Subleading contributions are given as a sum of
all available hydrodynamic vectors (not necessarily transverse) chosen in such
a way that the divergence of the current is non-negative when evaluated on the
solutions of equations of hydrodynamics. In the conformal case, up to 
second order in gradients, there are in total 5 available contributions
consisting  of the 3 hydrodynamic Weyl-invariant scalars \eqref{scalars}
multiplied by the 
velocity $u^{\mu}$ and 2 Weyl-invariant transverse vectors \eqref{vectors}
\beal{ecres}
J^\mu &=&  \frac{1}{4 \GN} b^{1-d} \left\{u^\mu  + 
b \left(j^\perp_1 V_{1}^{\mu} + j^\perp_2 V_{2}^{\mu} \right) + 
(j^{||}_1 S_{1}+j^{||}_2 S_{2}  + j^{||}_3 S_{3}) u^\mu \right\} \, .
\eea
The overall factor of $1/4 \GN$ in \eqref{ecres} comes from the holographic
expression for thermodynamic entropy. The on-shell divergence\footnote{In the
  sense of conservation of the energy-momentum tensor given by
  \eqref{enmomtensorUPtosecondORDER}} of the current \eqref{ecres} was
evaluated in reference \cite{Bhattacharyya:2008mz} and reads, up to third order
in gradients, 
\begin{equation}
\label{divergenceOFaCurrent}
\begin{split}
4\GN b^{d-1} \mathcal{D}_\mu J^\mu_S &= \frac{2b}{d}
\sigma^{\mu\nu}\left[ \sigma_{\mu\nu} -bd(d-2)\left(j^{\parallel}_3 -\frac{2
    (j^{\parallel}_2+j^{\perp}_2)}{d-2} \right)
  \omega_{\mu\lambda}\omega^\lambda{}_\nu\right.\\ 
&-bd(d-2)\left(j^{\parallel}_3 +
  \frac{1}{d(d-2)}\right)\left(\sigma_{\mu}{}^{\lambda}\sigma_{\lambda
    \nu}+u^\lambda \mathcal{D}_\lambda \sigma_{\mu\nu}+
  C_{\mu\alpha\nu\beta}u^\alpha u^\beta \right)\\ 
&\left.+\left((j^{\parallel}_1-j^{\perp}_1) bd+\tau_\omega\right) u^\lambda
  \mathcal{D}_\lambda \sigma_{\mu\nu} \right]\\ 
&+ b^2 (j^{\perp}_1+2 j^{\parallel}_3) \mathcal{D}_\mu \mathcal{D}_\nu
\sigma^{\mu\nu}+\ldots \\ 
\end{split}
\end{equation}
As understood in \cite{Romatschke:2009kr} for $d = 4$, this expression makes
it possible 
to constrain some of the coefficients appearing in \eqref{ecres}. These
arguments are based on the observation that local non-negativity should hold
both when the shear tensor vanishes at a given point, as well as when it is
arbitrary small (if it is large enough, then $\sigma_{\mu \nu} \sigma^{\mu
  \nu}$ dominates over other contributions and there are no further
constraints). The first condition automatically implies that 
\bel{condi1}
j^{\perp}_{1} = - 2 j^{\parallel}_{3},
\ee
whereas the second sets to zero all contributions which spoil non-negativity
for very small $\sigma_{\mu \nu}$, i.e.  
\bel{condi2}
j^{\parallel}_{2}+j^{\perp}_{2} = \frac{1}{2} \left( d - 2 \right)
j^{\parallel}_3  \quad \mathrm{and} \quad j^{\parallel}_3 = -
\frac{1}{d(d-2)}. 
\ee
Note that $j^{\perp}_{2}$ appears in the divergence only in the combination
$j^{\parallel}_{2}+j^{\perp}_{2}$, so that 
shifting $j^{\perp}_{2}$ and $j^{\parallel}_{2}$ keeping the sum constant does
not change the divergence. This ambiguity comes 
from the freedom of modifying the entropy current by adding a multiple of the 
divergence-free term $b^{3-d} {\cal D}_{\lambda} \omega^{\lambda \mu} =
b^{1-b} \left(- b V_{2}^{\mu} + S_{2} u^{\mu} \right)$
\cite{Bhattacharyya:2008mz,Bhattacharyya:2008xc,Booth:2010kr} and does not
affect the local rate of entropy production. At third order there are no
further 
constraints available so that $j^{\parallel}_{1}$ remains the only unspecified parameter
affecting the divergence \eqref{divergenceOFaCurrent}. Results of 
\cite{Bhattacharyya:2008xc,Bhattacharyya:2008mz} and \cite{Booth:2010kr} make
it clear that $j^{\parallel}_{1}$ is not fixed by some higher order argument
-- dual gravitational constructions, which all guarantee non-negativity of the
divergence,
lead to different values of $j^{\parallel}_{1}$. Thus, if the notion of local entropy
production in the near-equilibrium regime makes sense, there must be some
further constraints on 
the form of the hydrodynamic entropy current. This paper argues that one 
such constraint might be causality, which leads to considering the holographic
entropy current based on the apparent horizon in the dual gravity description. 

The problem of constructing a candidate hydrodynamic entropy current on the
gravity side of the correspondence was first solved in
\cite{Bhattacharyya:2008xc} and then generalized to weakly curved boundary
\cite{Bhattacharyya:2008ji} and to arbitrary dimensions
\cite{Bhattacharyya:2008mz}. These articles relied on using the bulk-boundary
map defined by ingoing null geodesics supplemented with boundary 
diffeomorphisms\footnote{The bulk-boundary map along ingoing null geodesics 
  associates points on the apparent horizon, the event horizon or any other
  ``generalized horizon'' with boundary points lying on the same null
  geodesics moving close to the boundary in a direction specified by a given
  vector field. This vector field is taken to be proportional to $u^{\mu}$ in
  the leading order with subleading corrections modifying dual entropy current
  at orders higher than 2. As anticipated in section \ref{quasilocalhorizons}, 
  in the gauge \eqref{wimetric} this bulk-boundary map acts trivially and maps
  points of the same $x^{\mu}$ position. Any such bulk-boundary map may be 
  supplemented with boundary diffeomorphisms, which are generated by another 
  vector field specified on the boundary. Such a vector field, if non-zero at 
  leading order of the gradient expansion, must be also proportional to 
  $u^{\mu}$, which modifies the dual entropy current at second and higher
  orders. The only parameter in \eqref{ecres} shifted by boundary
  diffeomorphisms of such form is $j^{\parallel}_{1}$. For a detailed
  discussion of bulk-boundary maps see \cite{Bhattacharyya:2008xc}.} 
to map the area form of the black brane event horizon satisfying the area theorem
onto a dual current of non-negative divergence. The main motivation for
mapping bulk data along ingoing null geodesics was causality. Note however
that such a constraint on the bulk-boundary map is self-consistent only when
the 
bulk entropy carrier is causal\footnote{Relaxing the assumption of causality
  of bulk-boundary maps has so far not been explored. Note at this point that
  although the mapping along ingoing null geodesics seems (at least
  superficially) to be causal, boundary diffeomorphisms composed with a given 
  bulk-boundary 
  map might lead to causality violations (see Section \ref{conclusions} for a
  discussion of this point).}. Such a notion is provided by an apparent
horizon, which along with the event horizon provides an example of a ``generalized
horizon'' introduced in \cite{Booth:2010kr}.  

The geometric setup described in Section \ref{quasilocalhorizons} contains a
distinguished vector field $v$ tangent to the horizon $\Delta$.  
As anticipated in \cite{Booth:2010kr} in the context of ``generalized
horizons'' one motivation for introducing $v$ is 
that the change of the area form on the horizon sections can be written in
terms of the expansion $\theta$ along $v$ 
\be
\tv = \frac{1}{\sqrt{h}} \mathcal{L}_{v} \sqrt{h} \, ,
\ee
where $h$ is the determinant of the induced metric on the section.
The generalized second law of thermodynamics is then the statement that the
area of the leaves is non-decreasing under the above flow 
\bel{ineq}
 \tv \geq 0 \, .
\ee
On the apparent horizon this area law is guaranteed by $\theta_{(\ell)} = 0$,
$\theta_{(n)} < 0$ and $C \geq 0$. The boundary entropy current is obtained from
$v$ by means of rewriting the left hand side of \eqref{ineq} within a chosen
bulk-boundary 
map as a divergence of boundary current. This current is interpreted as a
candidate boundary current and is given by \cite{Booth:2010kr} 
\bel{entrocur}
J^\mu = \frac{1}{4 \GN} \frac{1}{b} \sqrt{\frac{G}{g}}\ v^\mu \, .
\ee
where the prefactor involving $\GN$ has been introduced to reproduce
thermodynamic entropy at leading order and the AdS radius has been set to 1,
as in  \eqref{universalgravityaction}. The technical assumptions used to derive
\eqref{entrocur} match those in Section \ref{quasilocalhorizons}. In 
particular, the formula \eqref{entrocur} is valid for a trivial bulk-boundary
map, i.e. along null geodesics, which in the vicinity of the boundary move in
the direction defined by $u^{\mu}$. In the 
conformal case this direction can be modified 
only by second and higher order terms which change the entropy current at 
third order, and thus are beyond the scope of this article. The bulk-boundary
map used here is not 
supplemented with boundary diffeomorphisms partly due to causality reasons
(see Section \ref{conclusions} for more details). Because of this, the formula 
\eqref{entrocur} leads to the unique causal second order entropy current.

To apply \rf{entrocur} to the gravitational setup of Section
\eqref{quasilocalhorizons} one needs the form of $v$ and a 
computation of the determinant of the bulk metric $G$. One finds, up to
second order in gradients 
\bel{deter}
G = r^{2(d-1)} g\left(1 - K_{1} S_1 + \half \frac{1}{(br)^2} S_2 \right) \, .
\ee
As discussed earlier, the vector $v$ 
is completely fixed by the self-consistency of the  
bulk construction; the second order result \rf{vres} reads  
\bel{v2ord} 
v^{\mu} = b u^{\mu} + b^{2} ( \frac{2}{d \left( d -2
  \right)} V_1^{\mu} + \frac{1}{d-2} V_2^{\mu})\, .
\ee
The right hand side of \rf{entrocur} evaluated on the apparent
horizon \eqref{AH} leads to \rf{ecres} 
where $j^{\perp}_1$ and $j^{\perp}_2$ are fixed by Frobenius condition and equal
\bel{jperp}
j^\perp_1 = \frac{2}{d\left(d-2 \right)},  \quad j^\perp_2 =  \frac{1}{d-2}
\ee
while the $j^{||}_{i}$'s depend on the radial position of the apparent horizon
and read
\beal{ecahcoefs}
j^{||}_1 &=& - K_1(1) - \frac{1}{d} K_2(1) + \frac{2 \left(d-1\right)}{d \left( d - 2 \right)},
\nonumber \\  
j^{||}_2 &=&  \frac{\left(2-3 d \right)}{2 d \left(d-2 \right)}, \nonumber \\ 
j^{||}_3 &=& - \frac{1}{d \left(d-2 \right)} \, .
\eea
As a crosscheck one can easily see that the 
coefficients \eqref{ecahcoefs} satisfy conditions \eqref{condi1} and
\eqref{condi2}. Comparing 
with the event horizon result \cite{Bhattacharyya:2008mz,Booth:2010kr} one can
see that the only difference is in the choice of $j^{||}_{1}$ 
\be
j^{||}_{1, EH} = j^{||}_{1, AH}  - \frac{4}{d^2} \, .
\ee
Calculating the area theorem on the apparent horizon up to the second order in
gradients one obtains 
\be
\tl - C \tn\big |_{r=r_{AH}}  = \frac{2 S_1}{d} \, ,
\ee
which indeed matches the hydrodynamic result up to second order in 
gradients \eqref{divergenceOFaCurrent}. Calculating the third order contribution
in the bulk requires third order 
geometry, which has so far not been obtained. Nevertheless the match is
guaranteed by the 
formula \eqref{entrocur} relating divergence of the entropy current to the
area theorem on the apparent horizon modulo modifications of bulk-boundary
map.

\section{Summary\label{conclusions}}

This paper discusses the construction of apparent horizons in the geometry of
conformal fluid-gravity duality. The motivation for this work are interrelated
questions of local definition of entropy beyond equilibrium and foliation
dependence of apparent horizons of black holes. The reason for focusing on
apparent horizons is that they are causal objects: they evolve in response to
a flux of gravitational radiation or infall of matter. This makes them a
preferable carrier for the notion of entropy beyond equilibrium in dual
holographic field theories. There are however two caveats, which need to be
taken into account: foliation dependence of apparent horizons
\cite{Figueras:2009iu} and locality of entropy production being directly
related to the mapping of horizon information onto the boundary
\cite{Bhattacharyya:2008xc}. This paper focuses only on the first issue since
the latter appears most severely at higher orders of the gradient expansion than
are available for the geometry under consideration. 

The key idea behind this paper is that the apparent horizons of interest are only
those which are covariant in the sense of the dual hydrodynamic description,
i.e. can be covariantly specified (in the boundary sense) in terms of $b$,
$u^{\mu}$, $g_{\mu \nu}$ and their gradients. This requirement chooses only
those apparent horizons which have covariant dual hydrodynamic entropy
currents. Apparent horizons which do not satisfy this condition evade a
clear physical interpretation in terms of the dual field theory and are not
studied in this paper. Because of the requirement of hydrodynamic covariance, this paper 
adopts the somewhat unusual strategy of first finding suitable null normals and only 
afterwards confirming that they are foliation-forming, rather than starting with a foliation 
and then proceeding to normals. The approach adopted here relies on
the near-equilibrium regime, where one expects that at least one of the
apparent horizons will  ``closely'' follow the dynamics of the event horizon
\cite{Figueras:2009iu}. In the case of fluid-gravity duality this requirement
is indeed satisfied -- using even the results of \cite{Bhattacharyya:2008xc} alone one
can easily check that the leading order event horizon is at the same time an
(isolated) apparent horizon. It would certainly be interesting to try to apply similar
methods to find apparent horizons in other black hole spacetimes, perhaps
making contact with the framework of slowly evolving horizons
\cite{Booth:2003ji,Booth:2006bn,Kavanagh:2006qe}.

The main result of this paper is that, up to second order in gradients in
conformal fluid-gravity duality, there exists a unique apparent horizon
covariant in the hydrodynamic sense. It is very plausible that the uniqueness
of this apparent horizon holds to all orders of the gradient expansion, as
arguments in Section \ref{quasilocalhorizons} suggest. The apparent horizon in
the geometry of fluid-gravity duality is isolated at leading and first
subleading orders of the gradient expansion and becomes spatial once second
order gradient contributions are included. As expected
\cite{Bhattacharyya:2008xc} and confirmed by an explicit calculation in
\cite{Booth:2010kr}, the apparent horizon gives rise to a notion of
hydrodynamic entropy current when the area form on the apparent horizon is
mapped to the boundary in an appropriate way. Reference
\cite{Bhattacharyya:2008xc} introduced the map spanned along ingoing null
geodesics, the main motivation for it being the causal structure of bulk
spacetime. Such geodesics are specified by their tangent vector at the
boundary and hydrodynamic covariance forces this to be proportional to the
fluid velocity at leading order, but starting at second order additional
contributions appear. These terms will modify the form of the entropy current
at third and higher orders of the gradient expansion and are beyond the scope
of this paper. The only freedom, which affects the divergence of an entropy
current at second order comes from combining the bulk-boundary map specified
by ingoing bulk geodesics with boundary diffeomorphisms
\cite{Bhattacharyya:2008xc}. It is not clear however, whether or not this
leads to causality violations. One argument suggesting that it does was 
presented in \cite{Booth:2010kr}.  There, using the results from both
\cite{Bhattacharyya:2008xc,Bhattacharyya:2008mz} and the present paper, it was
shown that up to the second order in gradients the entropy current on the
event horizon is equivalent to the entropy current on the apparent horizon
when the bulk-boundary map in the latter case is supplemented with a 
particular boundary diffeomorphism. It would be very interesting to understand
better the constraints on the form of the bulk-boundary map which follow from
causality.

It seems unlikely that explicit causality violations due to the choice of the
bulk-boundary map can be visible at low orders of the gradient expansion in
the same way as in the near-equilibrium regime it is hard to tell whether the
event or one of apparent horizons is a better entropy carrier
\cite{Figueras:2009iu}. Instead, one probably needs to look for some general
principles or at concrete examples based on numerical solutions in
asymptotically AdS spacetimes. One such a simple example is the gravity dual
to boost-invariant flow
\cite{Janik:2005zt,Janik:2006ft,Heller:2007qt,Heller:2009zz,Kinoshita:2008dq,Kinoshita:2009dx,Booth:2009ct,Beuf:2009cx,Chesler:2009cy}. In
the boost-invariant case, the boundary dynamics depend on a single variable,
the proper time $\tau$, and its large proper time limit is governed by
boost-invariant hydrodynamics. Consider now the setup introduced in
\cite{Chesler:2009cy}, where the initial state given by the vacuum AdS space
is excited in a boost-invariant way in the vicinity of $\tau = 0$ by a
time-dependent boundary metric. Such a quench leads to the emission of
gravitational waves, and these propagate into the bulk and collapse forming a
black hole. The black hole equilibrates and in the end is well-described in
terms of fluid-gravity duality. In this example, the causal behavior of the
apparent horizon is clearly visible only in the vacuum (before the quench) and
in the far-from-equilibrium regime. However, close to equilibrium the apparent
and event horizon follow each other closely as expected, and from that
perspective there seems to be no reason to choose one over the other
\cite{Figueras:2009iu}. Consider now a slight modification of the setup
\cite{Chesler:2009cy}. As initial state at some late time $\tau_{i} \gg 0$ one
can choose a boost-invariant black brane solution dual to perfect fluid
hydrodynamics. Such a solution might have an arbitrary temperature and does
not produce any entropy as required by perfect fluid hydrodynamics. Consider
now the same kind of quench as considered in \cite{Chesler:2009cy}, but now in
the vicinity of $\tau_{i}$. Such a quench will excite both
far-from-equilibrium and hydrodynamics modes. The former equilibrate over a
time scale set by the inverse of temperature, so by taking temperature to be
large, one can effectively decouple them from analysis. Thus such setup serves
as a causally clear example of entropy production, governed entirely by
hydrodynamics. It would be very interesting to see 
what are the constraints on
the part of the bulk-boundary map which corresponds to boundary
diffeomorphisms following from this and 
similar examples.

As for more obvious further projects, it would be very interesting to
calculate the location of apparent horizons in the cases of charged
\cite{Banerjee:2008th,Erdmenger:2008rm}, non-conformal
\cite{Kanitscheider:2009as}, and superfluid
\cite{Bhattacharya:2011ee,Herzog:2011ec} fluid-gravity dualities.  In those
examples there are more gradient terms available so that the relevant
backgrounds may serve as further testing grounds for the claims of this 
paper. It would be also interesting to analyze the interplay between the
position of the event and apparent horizons in the context of the AdS/CFT
correspondence making use of the technology introduced in \cite{Booth:2010eu}.

In conclusion, the gravity dual to second order hydrodynamics of conformal media
in arbitrary dimensions has a unique apparent horizon, which is covariant in
the hydrodynamic sense. Possible ambiguities, which appear in the gravity
construction, should not affect the amount of entropy produced between two
equilibrium states (represented on the gravity side by two isolated
horizons). Furthermore if the foliation of the apparent horizon is fixed to
all orders in the gradient expansion and there are very stringent constraints on
the bulk-boundary map, then it is plausible that the local rate of entropy
production in the near-equilibrium regime is a meaningful observable. This
would be a very interesting result from the point of view of non-equilibrium
statistical mechanics.

\begin{acknowledgments}
The authors acknowledge the use of Kasper Peeters's excellent package {\tt
  Cadabra} \cite{DBLP:journals/corr/abs-cs-0608005,Peeters:2007wn}. This work
was partially supported by Polish Ministry of Science and Higher Education
grants \emph{N N202 173539} and \emph{N N202 105136}. MH acknowledges support from Foundation for Polish
Science. IB was supported by the Natural Sciences and Engineering Research
Council of Canada. This work is part of the research programme of the
Foundation for Fundamental Research on Matter, which is part of the
Netherlands Organisation for Scientific Research. 
\end{acknowledgments}

\bibliography{biblio}{}
\bibliographystyle{utphys}

\appendix
\section{Weyl-covariance and available gradient terms\label{appendixA}}
The Weyl-covariant derivative $\DD$ is defined so that for an arbitrary tensor
$Q^{\mu\ldots}_{\nu\ldots}$ of weight $w$ (i.e. one that obeys
$Q^{\mu\ldots}_{\nu\ldots} \rightarrow
e^{w\phi}Q^{\mu\ldots}_{\nu\ldots}$ 
under Weyl transformations
\eqref{Weylrescalings}),  $\mathcal{D}_\lambda\ Q^{\mu\ldots}_{\nu\ldots}$
transforms homogeneously with weight $w$ 
\begin{equation}\label{eqn:Ddef}
\begin{split}
\mathcal{D}_\lambda\ Q^{\mu\ldots}_{\nu\ldots} &\equiv
\nabla_\lambda\ Q^{\mu\ldots}_{\nu\ldots} + w\  \mathcal{A}_{\lambda}
Q^{\mu\ldots}_{\nu\ldots} \\ &+\brk{{g}_{\lambda\alpha}\mathcal{A}^{\mu} -
 \delta^{\mu}_{\lambda}\mathcal{A}_\alpha  -
\delta^{\mu}_{\alpha}\mathcal{A}_{\lambda}} Q^{\alpha\ldots}_{\nu\ldots} +
\ldots\\ 
&-\brk{{g}_{\lambda\nu}\mathcal{A}^{\alpha} -
  \delta^{\alpha}_{\lambda}\mathcal{A}_\nu  -
  \delta^{\alpha}_{\nu}\mathcal{A}_{\lambda}} Q^{\mu\ldots}_{\alpha\ldots} -
\ldots 
\end{split}
\end{equation}
The field $\mathcal{A}_{\mu}$ is given by \eqref{connectionFIELD} and
transforms as a connection under Weyl transformations
\eqref{connectionFIELD}. 

At first order in gradients there are two available 
contributions: $\DD_{\mu} u_{\nu}$ and $\DD_{\mu} b$. The latter quantity
vanishes at this order of the gradient expansion when
evaluated on solutions of the equations of perfect fluid hydrodynamics
\cite{Loganayagam:2008is} and  
thus at the first order the only nontrivial contribution comes from $\DD_{\mu}
u_{\nu}$. Taking its symmetric part one obtains the shear tensor $\sigma_{\mu
  \nu}$ 
\bel{shearintermsofDD}
\sigma_{\mu \nu} = \frac{1}{2} \DD_{(\mu} u_{\nu)},
\ee
whereas its antisymmetric part is the vorticity of the flow
\bel{vorticityintermsofDD}
\omega_{\mu \nu} = \frac{1}{2} \DD_{[\mu} u_{\nu]}.
\ee
Here $X_{(\mu \nu)} = X_{\mu \nu} +X_{\nu \mu}$ denotes symmetrization,
whereas $X_{[\mu \nu]} = X_{\mu \nu} - X_{\nu \mu}$ antisymmetrization. One
can show that both tensors \eqref{shearintermsofDD} and
\eqref{vorticityintermsofDD} are transverse, i.e. $u^{\mu} \sigma_{\mu \nu} =
u^{\mu} \omega_{\mu \nu} = 0$.  

Following \cite{Bhattacharyya:2008mz} one defines the Weyl-covariant Riemann
tensor $\cal{R}_{\mu \nu \lambda \sigma}$ 
\be
\mathcal{R}_{\mu\nu\lambda\sigma}= R_{\mu\nu\lambda\sigma}+ \nabla_{[\mu}
  \mathcal{A}_{\nu]} g_{\lambda\sigma} - 
\delta^\alpha_{[\mu}g_{\nu][\lambda}\delta^\beta_{\sigma]}\left(\nabla_\alpha
\mathcal{A}_\beta + \mathcal{A}_\alpha \mathcal{A}_\beta -
\frac{\mathcal{A}^2}{2} g_{\alpha\beta} \right)
\ee
Note that the Weyl-covariantized curvature tensors do not vanish even if the
fluid lives in a flat spacetime. With this, the  Weyl-covariant  Ricci tensor
$\mathcal{R}_{\mu\nu}$ and Weyl-covariant Ricci scalar $\mathcal{R}$ appearing
in \eqref{scalars} can be defined: 
\be
\mathcal{R}_{\mu\nu} = \mathcal{R}_{\mu\lambda\nu}{}^{\lambda} \quad \mathrm{and} \quad
\mathcal{R} = \mathcal{R}_{\lambda}{}^\lambda
\ee
Finally, the Weyl curvature tensor entering \eqref{tensors} is given by
\be
C_{\mu\nu\lambda\sigma} =
R_{\mu\nu\lambda\sigma} +
\frac{1}{d-2}\delta^\alpha_{[\mu}g_{\nu][\lambda}\delta^\beta_{\sigma]}
\left(\mathcal{R}_{\mu\nu} - \frac{\mathcal{R}g_{\mu\nu}}{2(d-1)}  
\right).
\ee

\section{Example: the Vaidya metric\label{appendixB}}
 
To see how the method for finding apparent horizon introduced in the context
of ``generalized horizons'' in \cite{Booth:2010kr} and reviewed in Section
\ref{quasilocalhorizons} works, consider the example of the Vaidya
metric\footnote{The letter $w$ denotes the ingoing Eddington-Finkelstein time
  coordinate to avoid clashing with the choice of $v$ for the vector which
  defines the slicing} 
\be
ds^2  =  2 dw dr - f(r,w)  dw^2 +  r^2 (d\theta^2 + \sin^2\theta r\phi^2)
\ee
with 
\be
f(r,w) = 1- \frac{2 m(w)}{r}.
\ee
In this case symmetry suggests taking $S = r - g(w)$ (i.e. spherically symmetric apparent horizon). Then the covector $m$ is 
\be
m_a = \left(dr - g'(w) dw\right)_a
\ee
and raising the index leads to
\be
m^a = \left(\partial_w + (f(r,w) - g'(w)) \partial_r \right)^a.
\ee
Here latin indices are used for $(r,w,\theta,\phi)$ and greek ones
(below) for $(w,\theta,\phi)$ -- this is sort of analogous to the conventions
used earlier for the AdS case. Choosing slicing on the horizon given by $v^\mu=(\partial_w)^\mu$ leads to 
\be
v^a = \left(\partial_w + g'(w)\partial_r\right)^a
\ee
Using \rf{evos} gives to 
\be
l = \partial_w + \half f(r,w)\ \partial_r  \quad \mathrm{and} \quad n =
\lambda  \left( f(r,w) - 2 g'(w)\right) \partial_r. 
\ee
Imposing the normalization condition \eqref{mvnormal} to fix $\lambda$ gives
the expected result \cite{Kavanagh:2006qe}  
\be
l =  \partial_w + \half f(r,w)\ \partial_r  \quad \mathrm{and} \quad n = - \partial_r.
\ee
Note that in spherical symmetry vectors $l$ and $n$ are fixed up to an overall scaling by conditions
\be
l_{a} l^{a} = n_{b} n^{b} = 0 \quad \mathrm{and} \quad l_{a} n^{a} = -1.
\ee
\end{document}